\documentclass[
    ,final            
    ,nonvarioref
  ]
  {aipproc}
\usepackage{amsmath}
\usepackage{amssymb}
\usepackage{amsfonts}
\usepackage{dcolumn}

\layoutstyle{8x11single}

\begin{document}

\newenvironment{ibox}[1]%
{\vskip 1.0em
\framebox[\columnwidth][r]{%
\begin{minipage}[c]{\columnwidth}%
\vspace{-1.0em}%
#1%
\end{minipage}}}
{\vskip 1.0em}

\newcommand{\iboxed}[1]{%
\vskip 1.0em
\framebox[\columnwidth][r]{%
\begin{minipage}[c]{\columnwidth}%
\vspace{-1.0em}
#1%
\end{minipage}}
\vskip 1.0em}

\newcommand{\fitbox}[2]{%
\vskip 1.0em
\begin{flushright}
\framebox[{#1}][r]{%
\begin{minipage}[c]{\columnwidth}%
\vspace{-1.0em}
#2%
\end{minipage}}
\end{flushright}
\vskip 1.0em}

\newcommand{\iboxeds}[1]{%
\vskip 1.0em
\begin{equation}
\fbox{%
\begin{minipage}[c]{1mm}%
\vspace{-1.0em}
#1%
\end{minipage}}
\end{equation}
\vskip 1.0em}

\def\Xint#1{\mathchoice
   {\XXint\displaystyle\textstyle{#1}}%
   {\XXint\textstyle\scriptstyle{#1}}%
   {\XXint\scriptstyle\scriptscriptstyle{#1}}%
   {\XXint\scriptscriptstyle\scriptscriptstyle{#1}}%
   \!\int}
\def\XXint#1#2#3{{\setbox0=\hbox{$#1{#2#3}{\int}$}
     \vcenter{\hbox{$#2#3$}}\kern-.5\wd0}}
\def\ddashint{\Xint=}
\def\dashint{\Xint-}

\newcommand{\alps}{\ensuremath{\alpha_s}}
\newcommand{\qbar}{\bar{q}}
\newcommand{\ubar}{\bar{u}}
\newcommand{\dbar}{\bar{d}}
\newcommand{\sbar}{\bar{s}}
\newcommand{\beq}{\begin{equation}}
\newcommand{\eeq}{\end{equation}}
\newcommand{\beqa}{\begin{eqnarray}}
\newcommand{\eeqa}{\end{eqnarray}}
\newcommand{\gs}{g_{\pi NN}}
\newcommand{\gw}{f_\pi}
\newcommand{\mq}{m_Q}
\newcommand{\mn}{m_N}
\newcommand{\mpi}{m_\pi}
\newcommand{\mrho}{m_\rho}
\newcommand{\momg}{m_\omega}
\newcommand{\bb}{\langle}
\newcommand{\kb}{\rangle}
\newcommand{\xvec}{\mathbf{x}}
\newcommand{\st}{\ensuremath{\sqrt{\sigma}}}
\newcommand{\Bvec}{\mathbf{B}}
\newcommand{\rvec}{\mathbf{r}}
\newcommand{\kvec}{\mathbf{k}}
\newcommand{\bvec}[1]{\ensuremath{\mathbf{#1}}}
\newcommand{\bra}[1]{\ensuremath{\bb#1|}}
\newcommand{\ket}[1]{\ensuremath{|#1\kb}}
\newcommand{\gft}{\ensuremath{\gamma_{FT}}}
\newcommand{\gfv}{\ensuremath{\gamma_5}}
\newcommand{\bfalp}{\ensuremath{\bm{\alpha}}}
\newcommand{\bfbeta}{\ensuremath{\bm{\beta}}}
\newcommand{\bfeps}{\ensuremath{\bm{\epsilon}}}
\newcommand{\lag}{{\lambda_\gamma}}
\newcommand{\lao}{{\lambda_\omega}}
\newcommand{\lN}{\lambda_N}
\newcommand{\lM}{\lambda_M}
\newcommand{\lB}{\lambda_B}
\newcommand{\epslag}{\ensuremath{\bm{\epsilon}_{\lag}}}
\newcommand{\bfept}{\ensuremath{\tilde{\bm{\epsilon}}}}
\newcommand{\bfgam}{\ensuremath{\bm{\gamma}}}
\newcommand{\bfnab}{\ensuremath{\bm{\nabla}}}
\newcommand{\bflambda}{\ensuremath{\bm{\lambda}}}
\newcommand{\bfmu}{\ensuremath{\bm{\mu}}}
\newcommand{\bfphi}{\ensuremath{\bm{\phi}}}
\newcommand{\bfvphi}{\ensuremath{\bm{\varphi}}}
\newcommand{\bfpi}{\ensuremath{\bm{\pi}}}
\newcommand{\bfsig}{\ensuremath{\bm{\sigma}}}
\newcommand{\bftau}{\ensuremath{\bm{\tau}}}
\newcommand{\bfrho}{\ensuremath{\bm{\rho}}}
\newcommand{\bfth}{\ensuremath{\bm{\theta}}}
\newcommand{\bfchi}{\ensuremath{\bm{\chi}}}
\newcommand{\bfxi}{\ensuremath{\bm{\xi}}}
\newcommand{\bfR}{\ensuremath{\bvec{R}}}
\newcommand{\bfP}{\ensuremath{\bvec{P}}}
\newcommand{\Rcm}{\ensuremath{\bvec{R}_{CM}}}
\newcommand{\spup}{\uparrow}
\newcommand{\spd}{\downarrow}
\newcommand{\up}{\uparrow}
\newcommand{\dn}{\downarrow}
\newcommand{\hbarom}{\frac{\hbar^2}{m_Q}}
\newcommand{\half}{\ensuremath{\frac{1}{2}}}
\newcommand{\thalf}{\ensuremath{\frac{3}{2}}}
\newcommand{\fhalf}{\ensuremath{\frac{5}{2}}}
\newcommand{\shalf}{\ensuremath{{\tfrac{1}{2}}}}
\newcommand{\sqtr}{\ensuremath{{\tfrac{1}{4}}}}
\newcommand{\sphalf}{\ensuremath{\genfrac{}{}{0pt}{1}{+}{}\!\tfrac{1}{2}}}
\newcommand{\smhalf}{\ensuremath{\genfrac{}{}{0pt}{1}{-}{}\!\tfrac{1}{2}}}
\newcommand{\sthalf}{\ensuremath{{\tfrac{3}{2}}}}
\newcommand{\spthalf}{\ensuremath{{\tfrac{+3}{2}}}}
\newcommand{\smthalf}{\ensuremath{{\tfrac{-3}{2}}}}
\newcommand{\sfhalf}{{\tfrac{5}{2}}}
\newcommand{\third}{{\frac{1}{3}}}
\newcommand{\tthird}{{\frac{2}{3}}}
\newcommand{\sthird}{{\tfrac{1}{3}}}
\newcommand{\stthird}{{\tfrac{2}{3}}}
\newcommand{\vnn}{\ensuremath{\hat{v}_{NN}}}
\newcommand{\vij}{\ensuremath{\hat{v}_{ij}}}
\newcommand{\vik}{\ensuremath{\hat{v}_{ik}}}
\newcommand{\argonne}{\ensuremath{v_{18}}}
\newcommand{\lqcd}{\ensuremath{\mathcal{L}_{QCD}}}
\newcommand{\lqed}{\ensuremath{\mathscr{L}_{QED}}}
\newcommand{\lgf}{\ensuremath{\mathcal{L}_g}}
\newcommand{\lqm}{\ensuremath{\mathcal{L}_q}}
\newcommand{\lqg}{\ensuremath{\mathcal{L}_{qg}}}
\newcommand{\nn}{\ensuremath{N\!N}}
\newcommand{\nnn}{\ensuremath{N\!N\!N}}
\newcommand{\qq}{\ensuremath{qq}}
\newcommand{\qqq}{\ensuremath{qqq}}
\newcommand{\qqb}{\ensuremath{q\bar{q}}}
\newcommand{\hpnd}{\ensuremath{H_{\pi N\Delta}}}
\newcommand{\hpqq}{\ensuremath{H_{\pi qq}}}
\newcommand{\hpqqa}{\ensuremath{H^{(a)}_{\pi qq}}}
\newcommand{\hpqqe}{\ensuremath{H^{(e)}_{\pi qq}}}
\newcommand{\hint}{\ensuremath{H_{\rm int}}}
\newcommand{\fpnn}{\ensuremath{f_{\pi\! N\!N}}}
\newcommand{\fenn}{\ensuremath{f_{\eta\! N\!N}}}
\newcommand{\gsnn}{\ensuremath{g_{\sigma\! N\!N}}}
\newcommand{\gpnn}{\ensuremath{g_{\pi\! N\!N}}}
\newcommand{\fpnd}{\ensuremath{f_{\pi\! N\!\Delta}}}
\newcommand{\grpg}{\ensuremath{g_{\rho\pi\gamma}}}
\newcommand{\gopg}{\ensuremath{g_{\omega\pi\gamma}}}
\newcommand{\fmqq}{\ensuremath{f_{M\! qq}}}
\newcommand{\gmqq}{\ensuremath{g_{M\! qq}}}
\newcommand{\fpqq}{\ensuremath{f_{\pi qq}}}
\newcommand{\gpqq}{\ensuremath{g_{\pi qq}}}
\newcommand{\feqq}{\ensuremath{f_{\eta qq}}}
\newcommand{\gonn}{\ensuremath{g_{\omega N\!N}}}
\newcommand{\gonna}{\ensuremath{g^t_{\omega N\!N}}}
\newcommand{\grnn}{\ensuremath{g_{\rho N\!N}}}
\newcommand{\gopr}{\ensuremath{g_{\omega\pi\rho}}}
\newcommand{\grnp}{\ensuremath{g_{\rho N\!\pi}}}
\newcommand{\grpp}{\ensuremath{g_{\rho\pi\pi}}}
\newcommand{\Lpnn}{\ensuremath{\Lambda_{\pi\! N\! N}}}
\newcommand{\Lonn}{\ensuremath{\Lambda_{\omega N\! N}}}
\newcommand{\Lonna}{\ensuremath{\Lambda^t_{\omega N\! N}}}
\newcommand{\Lrnn}{\ensuremath{\Lambda_{\rho N\! N}}}
\newcommand{\Lopr}{\ensuremath{\Lambda_{\omega\pi\rho}}}
\newcommand{\Lrpp}{\ensuremath{\Lambda_{\rho\pi\pi}}}
\newcommand{\getaqq}{\ensuremath{g_{\eta qq}}}
\newcommand{\fsqq}{\ensuremath{f_{\sigma qq}}}
\newcommand{\gsqq}{\ensuremath{g_{\sigma qq}}}
\newcommand{\piqq}{\ensuremath{{\pi\! qq}}}
\newcommand{\ylm}{\ensuremath{Y_\ell^m}}
\newcommand{\ylmc}{\ensuremath{Y_\ell^{m*}}}
\newcommand{\ebh}[1]{\hat{\bvec{e}}_{#1}}
\newcommand{\kbh}{\hat{\bvec{k}}}
\newcommand{\nbh}{\hat{\bvec{n}}}
\newcommand{\pvbh}{\hat{\bvec{p}}}
\newcommand{\qbh}{\hat{\bvec{q}}}
\newcommand{\Xbh}{\hat{\bvec{X}}}
\newcommand{\rbh}{\hat{\bvec{r}}}
\newcommand{\xbh}{\hat{\bvec{x}}}
\newcommand{\ybh}{\hat{\bvec{y}}}
\newcommand{\zbh}{\hat{\bvec{z}}}
\newcommand{\betabh}{\hat{\bfbeta}}
\newcommand{\tbh}{\hat{\bfth}}
\newcommand{\pbh}{\hat{\bfvphi}}
\newcommand{\dt}{\Delta\tau}
\newcommand{\kmag}{|\bvec{k}|}
\newcommand{\pmag}{|\bvec{p}|}
\newcommand{\qmag}{|\bvec{q}|}
\newcommand{\oas}{\ensuremath{\mathcal{O}(\alpha_s)}}
\newcommand{\vtxb}{\ensuremath{\Lambda_\mu(p',p)}}
\newcommand{\vtxp}{\ensuremath{\Lambda^\mu(p',p)}}
\newcommand{\pwqp}{e^{i\bvec{q}\cdot\bvec{r}}}
\newcommand{\pwqm}{e^{-i\bvec{q}\cdot\bvec{r}}}
\newcommand{\gsa}[1]{\ensuremath{\bb#1\kb_0}}
\newcommand{\oer}[1]{\mathcal{O}\left(\frac{1}{\qmag^{#1}}\right)}
\newcommand{\nub}[1]{\overline{\nu^{#1}}}
\newcommand{\epf}{E_\bvec{p}}
\newcommand{\epfp}{E_{\bvec{p}'}}
\newcommand{\eka}{E_{\alpha\kappa}}
\newcommand{\ekaq}{(E_{\alpha\kappa})^2}
\newcommand{\ekap}{E_{\alpha'\kappa}}
\newcommand{\ekpa}{E+{\alpha\kappa_+}}
\newcommand{\ekma}{E_{\alpha\kappa_-}}
\newcommand{\ekp}{E_{\kappa_+}}
\newcommand{\ekm}{E_{\kappa_-}}
\newcommand{\ekpap}{E_{\alpha'\kappa_+}}
\newcommand{\ekmap}{E_{\alpha'\kappa_-}}
\newcommand{\yjm}[1]{\mathcal{Y}_{jm}^{#1}}
\newcommand{\ysa}[3]{\mathcal{Y}_{#1,#2}^{#3}}
\newcommand{\yss}[2]{\mathcal{Y}_{#1}^{#2}}
\newcommand{\Dj}{\ensuremath{\mathscr{D}}}
\newcommand{\ysc}{\tilde{y}}
\newcommand{\enm}{\varepsilon_{NM}}
\newcommand{\Scg}[6]
	{\ensuremath{S^{#1}_{#4}\:\vphantom{S}^{#2}_{#5}
 	 \:\vphantom{S}^{#3}_{#6}\,}}
\newcommand{\Kmat}[6]
	{\ensuremath{K\left[\begin{array}{ccc} 
	#1 & #2 & #3 \\ #4 & #5 & #6\end{array}\right]}}
\newcommand{\irt}{\ensuremath{\frac{1}{\sqrt{2}}}}
\newcommand{\sirt}{\ensuremath{\tfrac{1}{\sqrt{2}}}}
\newcommand{\irth}{\ensuremath{\frac{1}{\sqrt{3}}}}
\newcommand{\sirth}{\ensuremath{\tfrac{1}{\sqrt{3}}}}
\newcommand{\irs}{\ensuremath{\frac{1}{\sqrt{6}}}}
\newcommand{\sirs}{\ensuremath{\tfrac{1}{\sqrt{6}}}}
\newcommand{\tors}{\ensuremath{\frac{2}{\sqrt{6}}}}
\newcommand{\stors}{\ensuremath{\tfrac{2}{\sqrt{6}}}}
\newcommand{\rtoth}{\ensuremath{\sqrt{\frac{2}{3}}}}
\newcommand{\rthot}{\ensuremath{\frac{\sqrt{3}}{2}}}
\newcommand{\ithrt}{\ensuremath{\frac{1}{3\sqrt{2}}}}
\newcommand{\Tg}{\ensuremath{\mathsf{T}}}
\newcommand{\irrep}[1]{\ensuremath{\mathbf{#1}}}
\newcommand{\cirrep}[1]{\ensuremath{\overline{\mathbf{#1}}}}
\newcommand{\Fij}{\ensuremath{\hat{F}_{ij}}}
\newcommand{\Fqij}{\ensuremath{\hat{F}^{(qq)}_{ij}}}
\newcommand{\Fsij}{\ensuremath{\hat{F}^{(qs)}_{ij}}}
\newcommand{\Opij}{\mathcal{O}^p_{ij}}
\newcommand{\fpij}{f_p(r_{ij})}
\newcommand{\titj}{\bftau_i\cdot\bftau_j}
\newcommand{\sisj}{\bfsig_i\cdot\bfsig_j}
\newcommand{\Sij}{S_{ij}}
\newcommand{\LS}{\bvec{L}_{ij}\cdot\bvec{S}_{ij}}
\newcommand{\TT}{\Tg_i\cdot\Tg_j}
\newcommand{\chet}{\ensuremath{\chi ET}}
\newcommand{\chpt}{\ensuremath{\chi PT}}
\newcommand{\chsy}{\ensuremath{\chi\mbox{symm}}}
\newcommand{\lchi}{\ensuremath{\Lambda_\chi}}
\newcommand{\lcon}{\ensuremath{\Lambda_{QCD}}}
\newcommand{\dcpsi}{\ensuremath{\bar{\psi}}}
\newcommand{\dc}[1]{\ensuremath{\overline{#1}}}
\newcommand{\dcpsip}{\ensuremath{\bar{\psi}^{(+)}}}
\newcommand{\psip}{\ensuremath{{\psi}^{(+)}}}
\newcommand{\dcpsim}{\ensuremath{\bar{\psi}^{(-)}}}
\newcommand{\psim}{\ensuremath{{\psi}^{(-)}}}
\newcommand{\llo}{\ensuremath{\mathcal{L}^{(0)}_{\chet}}}
\newcommand{\lchet}{\ensuremath{\mathcal{L}_{\chi}}}
\newcommand{\hchet}{\ensuremath{\mathcal{H}_{\chi}}}
\newcommand{\Hd}{\ensuremath{\mathcal{H}}}
\newcommand{\Dmu}{\ensuremath{\mathcal{D}_\mu}}
\newcommand{\Dsl}{\ensuremath{\slashed{\mathcal{D}}}}
\newcommand{\comm}[2]{\ensuremath{\left[#1,#2\right]}}
\newcommand{\acomm}[2]{\ensuremath{\left\{#1,#2\right\}}}
\newcommand{\ev}[1]{\ensuremath{\bb\hat{#1}\kb}}
\newcommand{\evt}[1]{\ensuremath{\bb{#1}(\tau)\kb}}
\newcommand{\evm}[1]{\ensuremath{\bb{#1}\kb_M}}
\newcommand{\evv}[1]{\ensuremath{\bb{#1}\kb_V}}
\newcommand{\ovl}[2]{\ensuremath{\bb{#1}|{#2}\kb}}
\newcommand{\pd}{\partial}
\newcommand{\pnpd}[2]{\frac{\partial{#1}}{\partial{#2}}}
\newcommand{\pppd}[1]{\frac{\partial{\hphantom{#1}}}{\partial{#1}}}
\newcommand{\plmu}{\partial_\mu}
\newcommand{\plnu}{\partial_\nu}
\newcommand{\pumu}{\partial^\mu}
\newcommand{\punu}{\partial^\nu}
\newcommand{\mcdf}{\delta^{(4)}(p_f-p_i-q)}
\newcommand{\ecdf}{\delta(E_f-E_i-\nu)}
\newcommand{\tr}{\mbox{Tr }}
\newcommand{\lxr}{\ensuremath{SU(2)_L\times SU(2)_R}}
\newcommand{\gV}[2]{\ensuremath{(\gamma^{-1})^{#1}_{\hphantom{#1}{#2}}}}
\newcommand{\gVd}[2]{\ensuremath{\gamma^{#1}_{\hphantom{#1}{#2}}}}
\newcommand{\LpV}[1]{\ensuremath{\Lambda^{#1}V}}
\newcommand{\hatH}{\ensuremath{\hat{H}}}
\newcommand{\hath}{\ensuremath{\hat{h}}}
\newcommand{\eht}{\ensuremath{e^{-\tau\hat{H}}}}
\newcommand{\ehdt}{\ensuremath{e^{-\Delta\tau\hat{H}}}}
\newcommand{\ehtm}{\ensuremath{e^{-\tau(\hat{H}-E_V)}}}
\newcommand{\ehdtm}{\ensuremath{e^{-\Delta\tau(\hat{H}-E_V)}}}
\newcommand{\Oop}{\ensuremath{\mathcal{O}}}
\newcommand{\Gop}{\ensuremath{\hat{\mathcal{G}}}}
\newcommand{\SU}[1]{\ensuremath{SU({#1})}}
\newcommand{\U}[1]{\ensuremath{U({#1})}}
\newcommand{\proj}[1]{\ensuremath{\ket{#1}\bra{#1}}}
\newcommand{\su}[1]{\ensuremath{\mathfrak{su}({#1})}}
\newcommand{\ip}[2]{\ensuremath{\bvec{#1}\cdot\bvec{#2}}}
\newcommand{\norm}[1]{\ensuremath{\left| #1\right|^2}}
\newcommand{\rnorm}[1]{\ensuremath{\lvert #1\rvert}}
\newcommand{\pid}{\left(\begin{array}{cc} 1 & 0 \\ 0 & 1\end{array}\right)}
\newcommand{\psx}{\left(\begin{array}{cc} 0 & 1 \\ 1 & 0\end{array}\right)}
\newcommand{\psy}{\left(\begin{array}{cc} 0 & -i \\ i & 0\end{array}\right)}
\newcommand{\psz}{\left(\begin{array}{cc} 1 & 0 \\ 0 & -1\end{array}\right)}
\newcommand{\ua}{\uparrow}
\newcommand{\da}{\downarrow}
\newcommand{\deln}{\delta_{i_1 i_2\ldots i_n}}
\newcommand{\GabRR}{G_{\alpha\beta}(\bfR,\bfR')}
\newcommand{\GRR}{G(\bfR,\bfR')}
\newcommand{\GfRR}{G_0(\bfR,\bfR')}
\newcommand{\GRiR}{G(\bfR_i,\bfR_{i-1})}
\newcommand{\GRRs}[2]{G(\bfR_{#1},\bfR_{#2})}
\newcommand{\Gdgn}{\Gamma_{\Delta,\gamma N}}
\newcommand{\Gdgnb}{\overline\Gamma_{\Delta,\gamma N}}
\newcommand{\GJT}{\Gamma_{LS}^{JT}(k)}
\newcommand{\GJTa}[2]{\Gamma^{#1}_{#2}}
\newcommand{\GtwJTa}[2]{\tilde{\Gamma}_{#1}^{#2}}
\newcommand{\Gtw}{\tilde{\Gamma}}
\newcommand{\Gbar}{\overline{\Gamma}}
\newcommand{\Gtil}{\tilde{\Gamma}}
\newcommand{\Gpndb}{\overline{\Gamma}_{\pi N,\Delta}}
\newcommand{\GbNgn}{{\overline{\Gamma}}_{N^*,\gamma N}}
\newcommand{\GNgn}{\Gamma_{N^*,\gamma N}}
\newcommand{\GbNmb}{{\overline{\Gamma}}_{N^*,MB}}
\newcommand{\Lg}[2]{\ensuremath{L^{#1}_{\hphantom{#1}{#2}}}}
\newcommand{\psik}{\ensuremath{\left(\begin{matrix}\psi_1 \\ \psi_2\end{matrix}\right)}}
\newcommand{\psib}{\ensuremath{\left(\begin{matrix}\psi^*_1&\psi^*_2\end{matrix}\right)}}
\newcommand{\Gf}{\ensuremath{\frac{1}{E-H_0}}}
\newcommand{\Gv}{\ensuremath{\frac{1}{E-H_0-\vnres}}}
\newcommand{\Gx}{\ensuremath{\frac{1}{E-H_0-V}}}
\newcommand{\Gex}{\ensuremath{\mathcal{G}}}
\newcommand{\Gfpm}{\ensuremath{\frac{1}{E-H_0\pm i\epsilon}}}
\newcommand{\vres}{v_R}
\newcommand{\vnres}{v}
\newcommand{\tpz}{\ensuremath{^3P_0}}
\newcommand{\tres}{t_R}
\newcommand{\tsr}{t^R}
\newcommand{\tsnr}{t^{NR}}
\newcommand{\trest}{\tilde{t}_R}
\newcommand{\tnres}{t}
\newcommand{\Pt}{P_{12}}
\newcommand{\Sz}{\ket{S_0}}
\newcommand{\Sa}{\ket{S^{(-1)}_1}}
\newcommand{\Sb}{\ket{S^{(0)}_1}}
\newcommand{\Sc}{\ket{S^{(+1)}_1}}
\newcommand{\sbasis}{\ket{s_1 s_2; m_1 m_2}}
\newcommand{\Sbasis}{\ket{s_1 s_2; S M}}
\newcommand{\sket}[2]{\ket{{#1}\,{#2}}}
\newcommand{\sbra}[2]{\bra{{#1}\,{#2}}}
\newcommand{\psmket}{\ket{\bvec{p};s\,m}}
\newcommand{\cket}{\ket{\bvec{p};s_1 s_2\,m_1 m_2}}
\newcommand{\hket}{\ket{\bvec{p};s_1 s_2\,\lambda_1\lambda_2}}
\newcommand{\hkets}{\ket{s\,\lambda}}
\newcommand{\phkets}{\ket{\bvec{p};s\,\lambda}}
\newcommand{\klsjm}{\ket{p;\ell s; j m}}
\newcommand{\pq}{\bvec{p}_q}
\newcommand{\pqb}{\bvec{p}_{\qbar}}
\newcommand{\mps}[1]{\frac{d^3{#1}}{(2\pi)^{3/2}}}
\newcommand{\mpsf}[1]{\frac{d^3{#1}}{(2\pi)^{3}}}
\newcommand{\du}[1]{u_{\bvec{#1},s}}
\newcommand{\dv}[1]{v_{\bvec{#1},s}}
\newcommand{\cdu}[1]{\overline{u}_{\bvec{#1},s}}
\newcommand{\cdv}[1]{\overline{v}_{\bvec{#1},s}}
\newcommand{\dus}[2]{u_{\bvec{#1},{#2}}}
\newcommand{\dvs}[2]{v_{\bvec{#1},{#2}}}
\newcommand{\cdus}[2]{\overline{u}_{\bvec{#1},{#2}}}
\newcommand{\cdvs}[2]{\overline{v}_{\bvec{#1},{#2}}}
\newcommand{\bop}[1]{b_{\bvec{#1},s}}
\newcommand{\dop}[1]{d_{\bvec{#1},s}}
\newcommand{\bops}[2]{b_{\bvec{#1},{#2}}}
\newcommand{\dops}[2]{d_{\bvec{#1},{#2}}}
\newcommand{\mev}{\mbox{ MeV}}
\newcommand{\gev}{\mbox{ GeV}}
\newcommand{\fmi}{\mbox{ fm}}
\newcommand{\M}{\mathcal{M}}
\newcommand{\Smat}{\mathcal{S}}
\newcommand{\JLSTh}{JLST\lambda}
\newcommand{\Tpg}{T_{\pi N,\gamma N}}
\newcommand{\tpg}{t_{\pi N,\gamma N}}
\newcommand{\vmbmb}{\ensuremath{v_{M'B',MB}}}
\newcommand{\tmbgn}{\ensuremath{t_{MB,\gamma N}}}
\newcommand{\Tonon}{\ensuremath{T_{\omega N,\omega N}}}
\newcommand{\tonon}{\ensuremath{t_{\omega N,\omega N}}}
\newcommand{\tronon}{\ensuremath{t^R_{\omega N,\omega N}}}
\newcommand{\Tonpn}{\ensuremath{T_{\omega N,\pi N}}}
\newcommand{\tonpn}{\ensuremath{t_{\omega N,\pi N}}}
\newcommand{\tronpn}{\ensuremath{t^R_{\omega N,\pi N}}}
\newcommand{\Tongn}{\ensuremath{T_{\omega N,\gamma N}}}
\newcommand{\tongn}{\ensuremath{t_{\omega N,\gamma N}}}
\newcommand{\trongn}{\ensuremath{t^R_{\omega N,\gamma N}}}
\newcommand{\vmbgn}{\ensuremath{v_{MB,\gamma N}}}
\newcommand{\vpngn}{\ensuremath{v_{\pi N,\gamma N}}}
\newcommand{\vongn}{\ensuremath{v_{\omega N,\gamma N}}}
\newcommand{\vonpn}{\ensuremath{v_{\omega N,\pi N}}}
\newcommand{\vpnpn}{\ensuremath{v_{\pi N,\pi N}}}
\newcommand{\vonon}{\ensuremath{v_{\omega N,\omega N}}}
\newcommand{\vrngn}{\ensuremath{v_{\rho N,\gamma N}}}
\newcommand{\tjtmbmb}{\ensuremath{t^{JT}_{M'B',MB}}}
\newcommand{\tjlsmngn}{\ensuremath{t^{JT}_{L'S'M'N',\lag\lN T_{N,z}}}}
\newcommand{\tjlsmbgn}{\ensuremath{t^{JT}_{LSMB,\lag \lN T_{N,z}}}}
\newcommand{\vjlsmngn}{\ensuremath{v^{JT}_{L'S'M'N',\lag \lN T_{N,z}}}}
\newcommand{\vjlsmbgn}{\ensuremath{v^{JT}_{LSMB,\lag \lN T_{N,z}}}}
\newcommand{\tjlsmnmb}{\ensuremath{t^{JT}_{L'S'M'N',LSMB}}}
\newcommand{\Tjlsmbmb}{\ensuremath{T^{JT}_{LSMB,L'S'M'B'}}}
\newcommand{\tjlsmbmb}{\ensuremath{t^{JT}_{LSMB,L'S'M'B'}}}
\newcommand{\tjlsmnpn}{\ensuremath{t^{JT}_{L'S'M'N',\ell \pi N}}}
\newcommand{\tjlsmbpn}{\ensuremath{t^{JT}_{LSMB,\ell \pi N}}}
\newcommand{\vjlsmnpn}{\ensuremath{v^{JT}_{L'S'M'N',\ell \pi N}}}
\newcommand{\vjlsmnmb}{\ensuremath{v^{JT}_{L'S'M'N',LSMB}}}
\newcommand{\vjlsmbpn}{\ensuremath{v^{JT}_{LSMB,\ell \pi N}}}
\newcommand{\Tjlsmngn}{\ensuremath{t^{R,JT}_{L'S'M'N',\lag\lN T_{N,z}}}}
\newcommand{\Tjlsmbgn}{\ensuremath{t^{R,JT}_{LSMB,\lag \lN T_{N,z}}}}
\newcommand{\Tfjlsmbgn}{\ensuremath{T^{JT}_{LSMB,\lag \lN T_{N,z}}}}
\newcommand{\Tjlsmnmb}{\ensuremath{t^{R,JT}_{L'S'M'N',LSMB}}}
\newcommand{\Tjlsmnpn}{\ensuremath{t^{R,JT}_{L'S'M'N',\ell \pi N}}}
\newcommand{\Tjlsmbpn}{\ensuremath{t^{R,JT}_{LSMB,\ell \pi N}}}
\newcommand{\Gbjlsi}{\ensuremath{{\Gamma}^{JT}_{LSMB,N^*_i}}}
\newcommand{\Gbjlspi}{\ensuremath{{\Gamma}^{JT}_{L'S'M'B',N^*_i}}}
\newcommand{\Gjlsi}{\ensuremath{\overline{\Gamma}^{JT}_{LSMB,N^*_i}}}
\newcommand{\Gijls}{\ensuremath{\overline{\Gamma}^{JT}_{N^*_i,LSMB}}}
\newcommand{\Gbijls}{\ensuremath{{\Gamma}^{JT}_{N^*_i,LSMB}}}
\newcommand{\Gjpn}{\ensuremath{\overline{\Gamma}^{JT}_{N^*_j,\ell\pn}}}
\newcommand{\Gign}{\ensuremath{\overline{\Gamma}^{JT}_{N^*_i,\lag\lN T_{N,z}}}}
\newcommand{\Gbign}{\ensuremath{{\Gamma}^{JT}_{N^*_i,\lag\lN T_{N,z}}}}
\newcommand{\Gjlsj}{\ensuremath{\overline{\Gamma}^{JT}_{LSMB,N^*_j}}}
\newcommand{\Gjem}{\ensuremath{\overline{\Gamma}^{JT}_{N^*_j,\lag\lN T_{N,z}}}}
\newcommand{\Ljtlsmbn}{\ensuremath{\Lambda^{JT}_{N^*LSMB}}}
\newcommand{\Drij}{\ensuremath{\mathcal{D}^{-1}_{ij}}}
\newcommand{\Mbres}{\ensuremath{M^{(0)}_{N^*}}}
\newcommand{\Cjtnlsmb}{\ensuremath{C^{JT}_{N^*LSMB}}}
\newcommand{\Ljtnlsmb}{\ensuremath{\Lambda^{JT}_{N^*LSMB}}}
\newcommand{\knstar}{\ensuremath{k_{N^*}}}
\newcommand{\vonen}{\ensuremath{v_{\omega N,\eta N}}}
\newcommand{\vonpd}{\ensuremath{v_{\omega N,\pi\Delta}}}
\newcommand{\vonsn}{\ensuremath{v_{\omega N,\sigma N}}}
\newcommand{\vonrn}{\ensuremath{v_{\omega N,\rho N}}}
\newcommand{\gnon}{\ensuremath{\gamma N\to \omega N}}
\newcommand{\gnpn}{\ensuremath{\gamma N\to \pi N}}
\newcommand{\gpop}{\ensuremath{\gamma p\to \omega p}}
\newcommand{\gppzp}{\ensuremath{\gamma p\to \pi^0 p}}
\newcommand{\gpppn}{\ensuremath{\gamma p\to \pi^+ n}}
\newcommand{\pnon}{\ensuremath{\pi N\to \omega N}}
\newcommand{\pnmb}{\ensuremath{\pi N\to MB}}
\newcommand{\gnmb}{\ensuremath{\gamma N\to M\!B}}
\newcommand{\onon}{\ensuremath{\omega N\to \omega N}}
\newcommand{\pmpon}{\ensuremath{\pi^- p\to \omega n}}
\newcommand{\pnpn}{\ensuremath{\pi N\to \pi N}}
\newcommand{\Gon}{\ensuremath{G_{0,\omega N}}}
\newcommand{\Gpn}{\ensuremath{G_{0,\pi N}}}
\newcommand{\rhomb}{\ensuremath{\rho_{MB}}}
\newcommand{\rhoon}{\ensuremath{\rho_{\omega N}}}
\newcommand{\rhopn}{\ensuremath{\rho_{\pi N}}}
\newcommand{\kon}{\ensuremath{k_{\omega N}}}
\newcommand{\kpn}{\ensuremath{k_{\pi N}}}
\newcommand{\Gmb}{\ensuremath{G_{0,MB}}}
\newcommand{\Tmbgn}{\ensuremath{T_{MB,\gamma N}}}
\newcommand{\vmbpgn}{\ensuremath{v_{M'B',\gamma N}}}
\newcommand{\pntpn}{\ensuremath{\pi N\!\to\!\pi N}}
\newcommand{\pnten}{\ensuremath{\pi N\!\to\!\eta N}}
\newcommand{\pnton}{\ensuremath{\pi N\!\to\!\omega N}}
\newcommand{\epos}{\ensuremath{\slashed{\epsilon}_{\lambda_\omega}}}
\newcommand{\epo}{\ensuremath{{\epsilon}_{\lambda_\omega}}}
\newcommand{\elevi}{\ensuremath{{\epsilon}_{\alpha\beta\gamma\delta}}}
\newcommand{\eps}{\ensuremath{\epsilon}}
\newcommand{\krho}{\ensuremath{\kappa_\rho}}
\newcommand{\komg}{\ensuremath{\kappa_\omega}}
\newcommand{\komga}{\ensuremath{\kappa^t_\omega}}
\newcommand{\doh}{\ensuremath{d^{(\half)}_{\lambda'\lambda}}}
\newcommand{\dohm}{\ensuremath{d^{(\half)}_{-\lambda,-\lambda'}}}
\newcommand{\dohmo}{\ensuremath{d^{(\half)}_{\lambda',-\half}}}
\newcommand{\dohpo}{\ensuremath{d^{(\half)}_{\lambda',+\half}}}
\newcommand{\Lor}[2]{\ensuremath{\Lambda^{#1}_{\hphantom{#1}{#2}}}}
\newcommand{\ILor}[2]{\ensuremath{\Lambda_{#1}^{\hphantom{#1}{#2}}}}
\newcommand{\LorT}[2]{\ensuremath{[\Lambda^T]^{#1}_{\hphantom{#1}{#2}}}}
\newcommand{\dsdo}{{\frac{d\sigma}{d\Omega}}}
\newcommand{\dspdo}{\ensuremath{{\frac{d\sigma_\pi}{d\Omega}}}}
\newcommand{\dsgdo}{\ensuremath{{\frac{d\sigma_\gamma}{d\Omega}}}}
\newcommand{\chipd}{\ensuremath{\chi^2/N_d}}
\newcommand{\chipda}{\ensuremath{\chi^2(\alpha)/N_d}}
\newcommand{\bpop}{\ensuremath{\bvec{p}'_1}}
\newcommand{\bptp}{\ensuremath{\bvec{p}'_2}}
\newcommand{\bpip}{\ensuremath{\bvec{p}'_i}}
\newcommand{\bpo}{\ensuremath{\bvec{p}_1}}
\newcommand{\bpt}{\ensuremath{\bvec{p}_2}}
\newcommand{\bpi}{\ensuremath{\bvec{p}_i}}
\newcommand{\bqo}{\ensuremath{\bvec{q}_1}}
\newcommand{\bqt}{\ensuremath{\bvec{q}_2}}
\newcommand{\bqi}{\ensuremath{\bvec{q}_i}}
\newcommand{\bQ}{\ensuremath{\bvec{Q}}}
\newcommand{\bq}{\ensuremath{\bvec{q}}}
\newcommand{\ketq}{\ensuremath{\ket{\bqo,\bqt}}}
\newcommand{\ketqc}{\ensuremath{\ket{\bQ,\bq}}}
\newcommand{\bP}{\ensuremath{\bvec{P}}}
\newcommand{\bPp}{\ensuremath{\bvec{P}'}}
\newcommand{\bpr}{\ensuremath{\bvec{p}}}
\newcommand{\bprp}{\ensuremath{\bvec{p}'}}
\newcommand{\ketPsiq}{\ensuremath{\ket{\Psi_{\bq}^{(\pm)}}}}
\newcommand{\ketPsiqQ}{\ensuremath{\ket{\Psi_{\bQ,\bq}^{(\pm)}}}}
\newcommand{\Ld}{\ensuremath{\mathcal{L}}}
\newcommand{\ps}{\mbox{ps}}
\newcommand{\fndp}{f_{N\Delta\pi}}
\newcommand{\fndr}{f_{N\Delta\rho}}
\newcommand{\said}{{\sc said}}
\newcommand{\ret}{\ensuremath{\langle{\tt ret}\rangle}}
\newcommand{\ddf}[1]{\ensuremath{\delta^{(#1)}}}
\newcommand{\Tpp}{\ensuremath{T_{\pi\pi}}}
\newcommand{\Kpp}{\ensuremath{K_{\pi\pi}}}
\newcommand{\Tpe}{\ensuremath{T_{\pi\eta}}}
\newcommand{\Kpe}{\ensuremath{K_{\pi\eta}}}
\newcommand{\Tep}{\ensuremath{T_{\eta\pi}}}
\newcommand{\Kep}{\ensuremath{K_{\eta\pi}}}
\newcommand{\Tee}{\ensuremath{T_{\eta\eta}}}
\newcommand{\Kee}{\ensuremath{K_{\eta\eta}}}
\newcommand{\Tpig}{\ensuremath{T_{\pi\gamma}}}
\newcommand{\Kpig}{\ensuremath{K_{\pi\gamma}}}
\newcommand{\oKpig}{\ensuremath{\overline{K}_{\pi\gamma}}}
\newcommand{\tKpig}{\ensuremath{\tilde{K}_{\pi\gamma}}}
\newcommand{\Teg}{\ensuremath{T_{\eta\gamma}}}
\newcommand{\Keg}{\ensuremath{K_{\eta\gamma}}}
\newcommand{\Kab}{\ensuremath{K_{\alpha\beta}}}
\newcommand{\R}{\ensuremath{\mathbb{R}}}
\newcommand{\C}{\ensuremath{\mathbb{C}}}
\newcommand{\Ezp}{\ensuremath{E^{\pi}_{0+}}}
\newcommand{\Eze}{\ensuremath{E^{\eta}_{0+}}}
\newcommand{\Ga}{\ensuremath{\Gamma_\alpha}}
\newcommand{\Gb}{\ensuremath{\Gamma_\beta}}
\newcommand{\RH}{\ensuremath{\mathcal{R}\!\!-\!\!\mathcal{H}}}
\newcommand{\calT}{\mathcal{T}}
\newcommand{\maid}{{\sc maid}}
\newcommand{\Kbar}{\ensuremath{\overline{K}}}
\newcommand{\zbar}{\ensuremath{\overline{z}}}
\newcommand{\kbar}{\ensuremath{\overline{k}}}
\newcommand{\dom}{\ensuremath{\mathcal{D}}}
\newcommand{\domi}[1]{\ensuremath{\mathcal{D}_{#1}}}
\newcommand{\pbar}{\ensuremath{\overline{p}}}
\newcommand{\Nab}{\ensuremath{N_{\alpha\beta}}}
\newcommand{\Nee}{\ensuremath{N_{\eta\eta}}}
\newcommand{\dth}[1]{\delta^{(3)}(#1)}
\newcommand{\dfo}[1]{\delta^{(4)}(#1)}
\newcommand{\intk}{\int\!\!\frac{d^3\! k}{(2\pi)^3}}
\newcommand{\intkg}{\int\!\!{d^3\! k_\gamma}}
\newcommand{\intks}{\int\!\!{d^3\! k_\sigma}}
\newcommand{\nch}{\ensuremath{N_{\mbox{ch}}}}
\newcommand{\nc}{\ensuremath{N_{ch}}}
\newcommand{\re}{\ensuremath{\mbox{Re }\!}}
\newcommand{\im}{\ensuremath{\mbox{Im }\!}}
\newcommand{\EetaS}{\ensuremath{E^\eta_{0+}}}
\newcommand{\EpiS}{\ensuremath{E^\pi_{0+}}}

\newcommand{\gn}{\ensuremath{\gamma N}}
\newcommand{\gp}{\ensuremath{\gamma p}}
\newcommand{\geta}{\ensuremath{\gamma \eta}}
\newcommand{\pp}{\ensuremath{pp}}
\newcommand{\pn}{\ensuremath{\pi N}}
\newcommand{\phn}{\ensuremath{\pi d}}
\newcommand{\en}{\ensuremath{\eta N}}
\newcommand{\epn}{\ensuremath{\eta' N}}
\newcommand{\pD}{\ensuremath{\pi \Delta}}
\newcommand{\sn}{\ensuremath{\sigma N}}
\newcommand{\rn}{\ensuremath{\rho N}}
\newcommand{\on}{\ensuremath{\omega N}}
\newcommand{\ppn}{\ensuremath{\pi\pi N}}
\newcommand{\kn}{\ensuremath{KN}}
\newcommand{\ky}{\ensuremath{KY}}
\newcommand{\kl}{\ensuremath{K\Lambda}}
\newcommand{\ks}{\ensuremath{K\Sigma}}
\newcommand{\bn}{\ensuremath{eN}}
\newcommand{\bpn}{\ensuremath{e\pi N}}
\newcommand{\fpo}{\ensuremath{5\oplus 1}}
\newcommand{\faoe}{{\sc FA08}}
\newcommand{\fpoe}{{\sc FP08}}
\newcommand{\fsoe}{{\sc FS08}}

\newcommand{\itPFP}{\textit{Physics for Future Presidents}}
\newcommand{\itaPFP}{\textit{PFP}}

\title{Toward a unified description of hadro- and photoproduction amplitudes}


\author{M.\ W.\ Paris}{
  address={Department of Physics, Data Analysis Center at the Center 
           for Nuclear Studies, The George Washington University,
           Washington, D.C. 20052, USA}
}

\author{R.\ A.\ Arndt}{
  address={Department of Physics, Data Analysis Center at the Center 
           for Nuclear Studies, The George Washington University,
           Washington, D.C. 20052, USA}
}

\author{R.\ L.\ Workman}{
  address={Department of Physics, Data Analysis Center at the Center 
           for Nuclear Studies, The George Washington University,
           Washington, D.C. 20052, USA}
}

\author{W.\ J.\ Briscoe}{
  address={Department of Physics, Data Analysis Center at the Center 
           for Nuclear Studies, The George Washington University,
           Washington, D.C. 20052, USA}
}

\author{I.\ I.\ Strakovsky}{
  address={Department of Physics, Data Analysis Center at the Center 
           for Nuclear Studies, The George Washington University,
           Washington, D.C. 20052, USA}
}

\begin{abstract}
The near-term objectives of the research program at the Data Analysis
Center are established within the context of the existing partial wave
analyses available through the online suite of analysis and database
codes accessible through \said, the Scattering Analysis Interactive
Database. This presentation reviews the efforts to determine a model
independent method to obtain sets of partial wave amplitudes for
strong and electromagnetic reactions, the interpretation of the
amplitudes in terms of the excited states of the nucleon, the role of
new precision unpolarized and polarized data, and new developments
aimed at determining the photoproduction mulitpoles in a unitary,
coupled-channel approach. The Chew-Mandelstam technique is discussed
and applied to the problem of the $S$-wave pion- and
eta-photoproduction amplitudes. The resulting eta production
amplitudes exhibit the expected resonant behavior near the eta
production threshold.  Application of this method to a unified
description of the hadro- and photoproduction amplitudes is discussed.
\end{abstract}

\maketitle


\section{Introduction}
\label{sec:intro}
Non-perturbative features of quantum chromodynamics are crucial in
determining the excitation spectrum of the nucleon. The model
independent determination of scattering and reaction amplitudes is a
necessary component to this end. A range of hadroproduction data,
including $\pn\to\pn$, $\pn\to\en$, $\pn\to\on$, and other inelastic
processes 
used to constrain theoretical models and phenomenological
parametrization of the scattering and reaction amplitudes.
Currently, however, a renaissance is underway in meson production and
resonance physics with reaction data issuing from a number of
precision electromagnetic facilities\cite{ Ajaka:1998zi, Crede:2003ax,
Nakabayashi:2006ut, Elsner:2007hm, McGeorge:2007tg, Dugger:2009pn,
Williams:2009yj}.  Phenomenological efforts to analyze these data
consistent with some subset of constraints imposed by quantum field
theory are under current study. The quality and quantity of data in
electromagnetic induced reactions is becoming sufficient to rival and
possibly surpass the hadroproduction data.  Since the electromagnetic
reactions proceed mainly through the hadronic channels, the new data
offers the possibility of ``back-constraining'' the hadronic
amplitudes, conventionally determined only in fits to the
hadroproduction data.

This presentation describes an exploratory study of the $S$--wave $\pi$--
and $\eta$--photoproduction multipoles in the ``Chew-Mandelstam''
approach, related to the $N/D$ representation, to the electromagnetic
reaction amplitude. The novel concept, which provokes and permits this
exploratory study, is the generalization of the Chew-Mandelstam
approach to the electromagnetic sector. We have developed a new form
for the amplitude that incorporates multichannel hadronic rescattering
effects in a complete manner consistent with
unitarity\cite{Paris:2010tz}.  The near-term
objective is to develop a framework in which to analyze the hadro- and
electroproduction reactions simultaneously in a global framework.


\section{Chew-Mandelstam parametrization}
\label{sec:formal}
Previous work in the determination of the $\eta$ photoproduction
amplitudes\cite{Green:1997yia,Arndt:1998nm} has shown 
that an approach
which includes the coupling of the electromagnetic channel to the
$\pn$ and $\en$ channels in the region of energies near the center-of-mass
energy, $W=1535$ MeV gives a reasonably good description of the data and
a plausible form for the amplitudes. However, as our ultimate
objective is the simultaneous parametrization of hadro- and 
photoproduction scattering and reaction observables, we will go beyond
the two-channel treatment for this study of the $\Eze$ multipole amplitude.

The form of the Chew-Mandelstam parametrization, which we employ
in this study follows as a consequence of the analytic structure 
imposed by the 
unitarity\cite{Zimmerman:1961aa} 
of the $S$ matrix in the physical region, $W>m_i+m_t$, 
where $W$ is the center-of-mass energy and $m_i$ and $m_t$ are
the masses of the incident and target particles. 
Confining our attention to two-particle initial and final states,
the $S$ matrix is 
%
unitary $S^\dag S = S S^\dag = 1.$
In the partial wave representation with $S=1+2i\tilde\rho T = 1+ 2i T'$ we have
\begin{align}
\label{eqn:uct-thr}
\mbox{Im } T'^{-1} &= -\theta(W-M_+),
\end{align}
where $M_+$ is the diagonal matrix of threshold openings. 
Since this equation isolates the imaginary part
of the inverse-$T$ matrix, we may write
\begin{align}
T'^{-1} &= \mbox{Re } T'^{-1} + i\mbox{Im } T'^{-1}, \\
\label{eqn:Kinv}
       &= K'^{-1} - i\theta(W-M_+),
\end{align} 
where we've defined $\mbox{Re } T'^{-1} = K'^{-1}$.
Multiplying from one
side by $T'$ and the other by $K'$ gives the Heitler integral
equation\cite{Heitler:1941a}
\begin{align}
\label{eqn:TpK}
T' &= K' + K' i\theta(W-M_+) T'.
\end{align}
This is the starting point for the Chew-Mandelstam parametrization of
the reaction amplitude.
Equation \eqref{eqn:uct-thr} implies discontinuities
in the derivative of the imaginary part at each channel threshold
$W=m_{\sigma +}$.
The violation of the Cauchy-Riemann equations at threshold indicates
the presence of a branch point. 

The partial wave amplitude
has the following singularities.
There are branch points in the physical region at the channel-opening 
thresholds as in Eq.\eqref{eqn:uct-thr}, branch points in the region $W<0$,
and possible poles consistent with causality\cite{Eden:1966sm}. 
An efficient parametrization following Ref.\cite{Chew:1960iv}, which 
encodes these singularities, involves the factorization of the partial 
wave amplitude. This is referred to as the ``$N/D$'' approach. We will 
use the $N/D$ language to clarify the nature of the singularities of the 
$T$ matrix which are included and those neglected in our Chew-Mandelstam 
approach.

The preceding discussion of unitarity and the $N/D$ approach provides
the context for our present parametrization.
The Chew-Mandelstam parametrization developed here is similar to 
those of Refs.\cite{Basdevant:1978tx}
and \cite{Arndt:1985vj}. 
We consider Eq.\eqref{eqn:Kinv} and rewrite it, confining our attention
to the $S$ wave multipole as
\begin{align}
\label{eqn:TinvK}
T^{-1} &= K^{-1} - i \tilde\rho \\
&= (K^{-1} + \mbox{Re}\,C) - (\mbox{Re}\,C + i\tilde\rho) \nonumber \\
\label{eqn:TinvKbar}
 &= \Kbar^{-1} - C,
\end{align}
where $\tilde\rho = \rho\theta(W-M_+)$ and $\mbox{Im}\,C = \tilde\rho
= \theta(W-M_+)\rho$. The transition matrix is given in terms of the
``Chew-Mandelstam'' (CM) $K$ matrix, \Kbar\ by
\begin{align}
\label{eqn:TKbar}
T &= \Kbar + \Kbar C T.
\end{align}



Equation \eqref{eqn:TKbar} fixes our Chew-Mandelstam parametrization. In
the language of the $N/D$ approach, we have neglected the $W<0$ branch
points of $N$ and made the approximation $N(W)=\Kbar(W)$, an entire function. 
The ``Chew-Mandelstam'' function, $C_\alpha$ is determined solely by the 
unitarity constraint, Eq.\eqref{eqn:uct-thr} 
The Chew-Mandelstam function
is given by a Cauchy integral over the discontinuity of $C_\alpha$ in
the physical region with a single subtraction.

The relationship between the Heitler
$K$ matrix and the CM $K$ matrix, $\Kbar$ is given by
\begin{align}
\label{eqn:KKbar}
K &= \Kbar + \Kbar [\mbox{Re}\, C] K.
\end{align}
This demonstrates a possible advantage of using the CM $K$ matrix. If
we consider a polynomial parametrization of a given CM $K$ matrix element
then we see, by solving Eq.\eqref{eqn:KKbar} for $K$
\begin{align}
K &= \frac{1}{1-\Kbar [\mbox{Re}\,C]} \Kbar
\end{align}
that poles may appear in the $K$
matrix. 
Attempts to relate the $K$ matrix poles to resonances have been
made\cite{Ceci:2006jj}. 
Here, we simply 
point out that, though $K$ matrix poles are not simply related to
$T$ matrix poles\cite{Workman:2008iv}, Eq.\eqref{eqn:KKbar} shows that
one need not explicitly include pole terms in $\Kbar$ in order
to have poles in $K$.

\section{Results}
\label{sec:results}
The Chew-Mandelstam parametrization for the $T$ matrix,
described in the preceding section, has been applied
recently\cite{Arndt:2006bf} to a coupled-channel 
fit for the $\pn$ elastic scattering and $\pn\to\en$ reaction.
It gives a realistic description of the data with $\chi^2$ per
datum better than any other parametrization or model, to our
knowledge. 
The current \said\ parametrization 
used in this fit is given as
\begin{align}
\label{eqn:Tab}
T_{\alpha\beta} &= \sum_\sigma [1-\Kbar C]^{-1}_{\alpha\sigma}
\Kbar_{\sigma\beta}
\end{align}
where $\alpha,\beta$ and $\sigma$ are channel indices for the 
considered channels,
$\pn,\pD,\rn$ and $\en$. 
Given the success of this approach
in the hadronic two-body sector, the application to the study
of meson photoproduction is warranted.

The central result of the current exploratory study is to show that this
form can be extended to include the electromagnetic channel, 
\begin{align}
\label{eqn:Tag}
T_{\alpha\gamma} &= \sum_\sigma [1-\Kbar C]^{-1}_{\alpha\sigma}
\Kbar_{\sigma\gamma}
\end{align}
where $\gamma$ denotes the electromagnetic channel, $\gn$. Note that
Eqs.\eqref{eqn:Tab} and \eqref{eqn:Tag} share the common factor,
$[1-\Kbar C]_{\alpha\sigma}^{-1}$ which encodes, at least 
qualitatively speaking, the hadronic channel coupling 
(or rescattering) effects.

\begin{figure}[t]
\includegraphics[width=250pt,keepaspectratio,clip]{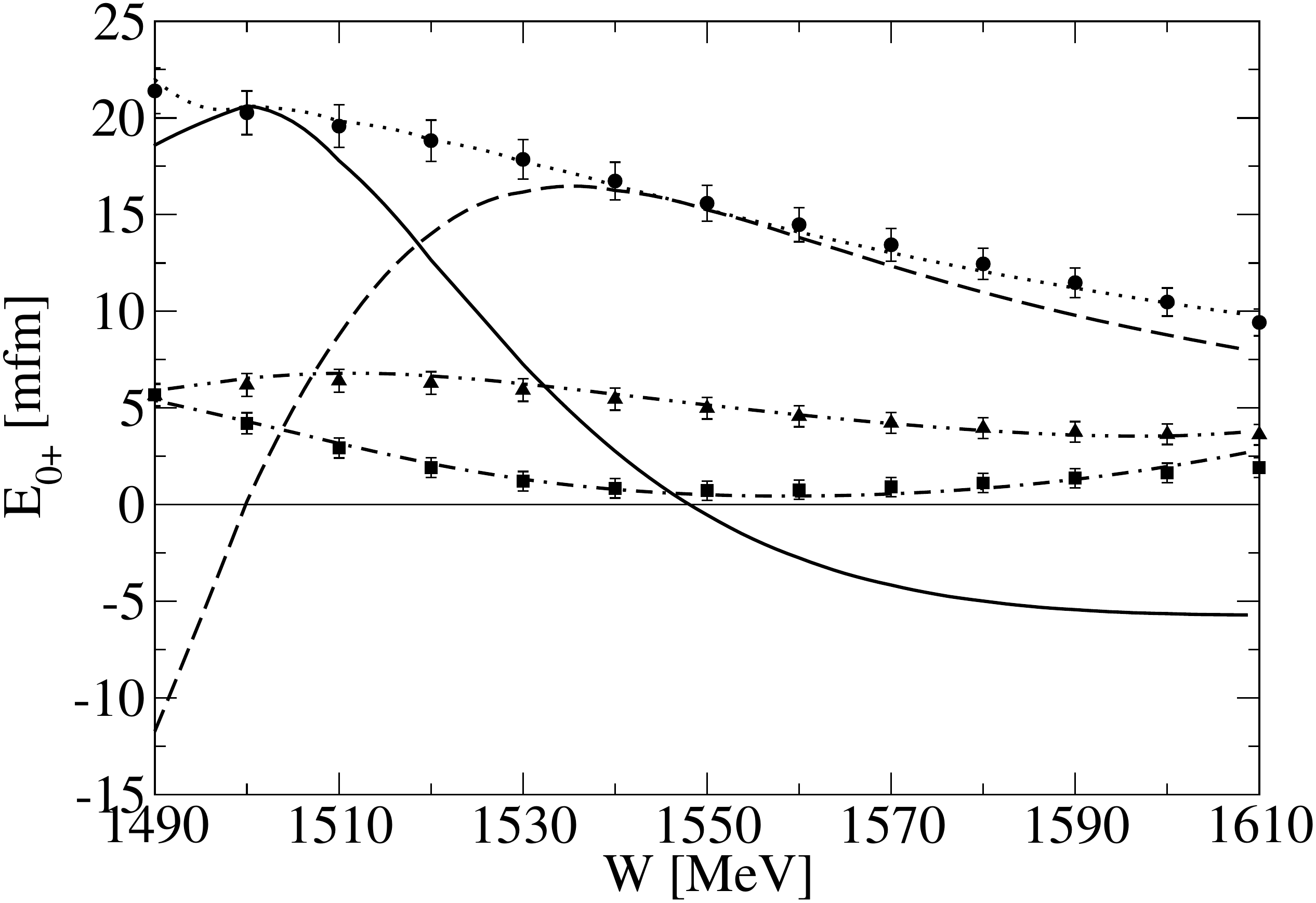}
\caption{\label{fig:Esaid}The predicted values for the real (solid curve) and
imaginary (dashed curve) for \EetaS versus the energy, $W$. 
The modulus $|\EetaS|$ (dotted curve),
the real (dot-dashed curve), and the imaginary (double
dot-dashed curve) parts of the $\pi$--photoproduction, $\EpiS$ were fit
to pseudodata generated from the \said\ solution\cite{Arndt:1989ww} with
the parametrized form Eq.\eqref{eqn:Tag} using 8 parameters.}
\end{figure}

The form Eq.\eqref{eqn:Tag} for photoproduction should be contrasted 
with that currently employed in the $\pi$--photoproduction studies of 
Refs.\cite{Arndt:1989ww,Arndt:1998nm}
\begin{align}
\label{eqn:Tgamma-old}
T_{\pi\gamma} &= A(W) (1+iT_{\pi\pi}(W)) + B(W)T_{\pi\pi}(W)
\end{align}
where the ``structure functions'' $A(W)$ and $B(W)$ are parametrized
as polynomials in the energy, $W$, $T_{\pi\gamma}
=T_{\pn,\gn}$ and $T_{\pi\pi}=T_{\pn,\pn}$, and the factor $A(W)$
contains a contribution from tree-level Born diagrams. This satisfies
Watson's theorem\cite{Watson:1952ji} (as does Eq.\eqref{eqn:Tag}), 
and is derived via the considerations
discussed in Ref.\cite{Workman:2005eu}.
While resulting in a realistic description of the data and being
comparable, at least qualitatively, with other parametrization such
as \maid\cite{Chiang:2001as} for $\pi$--photoproduction, 
it does not satisfy 
the full multichannel unitarity constraint imposed by Eq.\eqref{eqn:uct-thr}.
This deficiency led us to consider the form in Eq.\eqref{eqn:Tag}, which 
manifestly satisfies the multichannel unitarity constraint, 
Eq.\eqref{eqn:uct-thr}.

In the present 
study we perform a coupled-channel fit of the modulus $|\EetaS(W)|$ 
and (the real and imaginary parts of) the existing \said\ and
\maid\ $\pi$--photoproduction amplitudes,
$\EpiS$ in the $S_{11}(1535)$ resonance region. (See
Ref.\cite{Paris:2010tz} for comparisons to \maid.)
The fit was carried
out by taking the factor $[1-\Kbar C]_{\alpha\sigma}^{-1}$ 
in Eq.\eqref{eqn:Tag} as determined in the
the hadronic study of Ref.\cite{Arndt:2006bf} and adjusting the
parameters of $\Kbar_{\sigma\gamma}$ 
(discussed in detail in the subsections below).
The 
phase of the \EetaS\ multipole in this study gives a resonant wave and
encourages us to continue with this approach. 
Figure \eqref{fig:Esaid} shows the result of fitting the
modulus $|\EetaS|$ and the real and imaginary parts of the
\said\ {\sc sp09} \EpiS\ multipole \cite{Dugger:2009pn}
using an eight-parameter fit. 
The Chew-Mandelstam \Kbar\ matrix was assumed to have 
four channels, $\pn,\pD,\rn$, and $\en$.
Eight parameters were varied in the fit to a total of 
113 pseudodata points including the modulus $|\EetaS|$
over the energy range 1490 MeV $\le W \le $ 1610 MeV and the amplitude 
$\EpiS$ over the energy range 1120 MeV $\le W \le $1610. The $\chi^2$
per datum over for the fits to the pseudodata, generated with the
\said\ interactive code facility
were less than one
in all of the fits made in this work including those in the region
1120 MeV $\le W \le$ 1490 MeV which are not displayed in order to keep
the figures manageable and focus attention on the $S_{11}(1535)$ resonance
region.

%

\section{Conclusion and ongoing work}
\label{sec:conclusion}
We reviewed the implication of unitarity on the analytic structure of
the single meson production scattering and reaction amplitudes.
We related the Chew-Mandelstam $K$-matrix parametrization
to the $N/D$ approach, showing that the parametrization of the $\Kbar$
matrix neglects the effects of the distant left-hand cut. 
Using the Chew-Mandelstam $K$ matrix $\Kbar$, we performed a simultaneous
coupled-channel fit of the $\eta$--photoproduction $S_{11}$ multipole 
modulus, $|\EetaS|$ and the $\pi$--photoproduction amplitude, $\EpiS$.
The parametrization was restricted only to the CM $K$ matrix elements
$\Kbar_{\sigma\gamma}$ in Eq.\eqref{eqn:Tag}, while the $[1-\Kbar C]^{-1}$
factors were taken from the existing \said\ fits to the hadronic data. 
The anticipated resonant structure for the phase of the \EetaS\ multipole
was demonstrated in fits to \said\ amplitudes.

The results of the exploratory study indicate that this is a reasonable
approach toward the objective of
determining a complete set of scattering and reaction amplitudes for
$\pn\to\pn$, $\pn\to\en$, $\gn\to\pn$, and $\gn\to\en$ processes in
a multichannel unitary formalism.
The ultimate objective of the study will be a simultaneous fit to both the
hadronic and electromagnetic scattering and reaction observables and will
constitute, at least for two-body unitarity, a global description of the
hadro- and photoproduction amplitudes.


\begin{theacknowledgments}
This work is dedicated to the memory of R.\ Arndt without whom it
would not have been possible; it was supported by the US Department of
Energy Grant No.\ DE-FG02-99ER41110.
\end{theacknowledgments}



\bibliographystyle{aipproc}   

\bibliography{master}

\IfFileExists{\jobname.bbl}{}
 {\typeout{}
  \typeout{******************************************}
  \typeout{** Please run "bibtex \jobname" to optain}
  \typeout{** the bibliography and then re-run LaTeX}
  \typeout{** twice to fix the references!}
  \typeout{******************************************}
  \typeout{}
 }

\end{document}